**Demonstration of Circadian Rhythm in Heart Rate Turbulence**

**using Novel Application of Correlator Functions**

Mari A Watanabe, MD, PhD[1,2], Mark Alford, PhD[3], Raphael Schneider[4], Axel Bauer, MD[4],

Petra Barthel, MD[4], Phyllis Stein, PhD[5], Georg Schmidt, MD[4]

[1] Dept. of Internal Medicine, St. Louis University, 3635 Vista Avenue, St. Louis, MO, 63110, USA

[2] Institute of Biomedical Life Sciences, Glasgow University, Glasgow, G12 8QQ, UK

[3] Dept. of Physics, Washington University, 1 Brookings Drive, St. Louis, MO, 63130, USA

[4] 1. Medizinische Klinik der Technischen Universität München, Ismaninger Strasse 22, München, 81675, Germany

[5] Heart Rate Variability Lab, Washington University, 4625 Lindell Blvd., St. Louis, MO 63108, USA

**short title**: Correlator Functions for Assessing Circadian Rhythms

**support**: Mari Watanabe was supported by British Heart Foundation Project Grant PG/02/155 during the initial phases of this research.

**conflict of interest**: Georg Schmidt is the holder of a patent on the heart rate turbulence algorithm

7100 words, 36 references, 4 figures

**Correspondence to:**

Mari A Watanabe, MD, PhD
Dept. of Internal Medicine
St. Louis Univ. School of Medicine
3635 Vista Avenue
St. Louis, MO 63110
tel : 314-577-8860
fax: 314-577-8861
watanabe@slu.edu

**Abstract**

**Background**: It has been difficult to demonstrate circadian rhythm in the two parameters of heart rate turbulence, turbulence onset (TO) and turbulence slope (TS).

**Objective**: To devise a new method for detecting circadian rhythm in noisy data, and apply it to selected Holter recordings from two post-myocardial infarction databases, Cardiac Arrhythmia Suppression Trial (CAST, n=684) and Innovative Stratification of Arrhythmic Risk (ISAR, n=327).

**Methods**: For each patient, TS and TO were calculated for each hour with >4 VPCs. An autocorrelation function Corr($\Delta$t) = <TS(t) TS(t+$\Delta$t)> was then calculated, and averaged over all patients. Positive Corr($\Delta$t) indicates that TS at a given hour and $\Delta$t hours later are similar. TO was treated likewise. Simulations and mathematical analysis showed that circadian rhythm required Corr($\Delta$t) to have a U-shape consisting of positive values near $\Delta$t=0 and 23, and negative values for intermediate $\Delta$t. Significant deviation of Corr($\Delta$t) from the correlator function of pure noise was evaluated as a chi-squared value.

**Results**: Circadian patterns were not apparent in hourly averages of TS and TO plotted against clock time, which had large error bars. Their correlator functions, however, produced chi-squared values of ~10 in CAST (both p<0.0001) and ~3 in ISAR (both p<0.0001), indicating presence of circadian rhythmicity.

**Conclusion**: Correlator functions may be a powerful tool for detecting presence of circadian rhythms in noisy data, even with recordings limited to 24 hours.

**Abbreviations**:

HRT: heart rate turbulence

TO: turbulence onset

TS: turbulence slope

VPC: ventricular premature contraction

CAST: Cardiac Arrhythmia Suppression Trial

ISAR: Innovative Stratification of Arrhythmic Risk trial

**Introduction**

The goal of our study was to demonstrate presence of a circadian rhythm in heart rate turbulence (HRT), a risk stratifier for mortality in patients after myocardial infarction (1-3). The phenomenon refers to the baroreflex-based oscillation in heart rate seen immediately following a ventricular premature contraction (VPC), namely, a brief acceleration followed by a deceleration that overshoots baseline, then returns to baseline within approximately 20 beats. Two parameters measured from 24 hour electrocardiographic (Holter) records are used to characterize HRT: turbulence onset (TO) and turbulence slope (TS), which quantify the heart rate acceleration and deceleration, respectively. Myocardial infarction patients with reduced HRT (TO>0%, TS<2.5 ms per beat) are at high risk for death (1,2) and sudden cardiac arrest (4,5). We wanted to demonstrate a circadian rhythm for two reasons. One was to determine the possibility of assessing a patient's HRT from electrocardiograms recorded during a selected time of day, rather than over the conventional 24 hours. The second was to confirm the influence of autonomic tone on HRT. During sleep, there is generally sympathetic withdrawal and increase of vagal tone. If HRT is predominantly under vagal control as a study using atropine suggests (6), then HRT should be greater during sleep.

HRT relies on presence of VPCs for measurement. Therefore, we could not use young healthy volunteers whose autonomic function is better for our study, because they have too few spontaneous VPCs to allow a study of circadian pattern of HRT. However, demonstrating a circadian rhythm in HRT using data from patients with prior myocardial infarction using conventional methods as we did here, also proved to be difficult. Circadian patterns in HRT were equivocal because of large error bars. A major reason for this is the commonly seen uneven distribution of VPC occurrence over the day, which contributes doubly to noise: because of gaps in some hours in some patients, and because TS is known to decrease in value as greater numbers of VPCs are averaged for its calculation (7).

Studies in the literature commonly try to demonstrate presence of circadian rhythm by showing a plot of the hourly population average of a variable over 24 hours (hereafter referred to as a pattern plot) and rely on visual inspection to persuade the reader that a pattern exists, or by selecting 2 or more segments from the 24 hours and comparing their values statistically. The former method is qualitative and is not persuasive when the data is noisy. The latter technique is open to abuse, because one has the freedom to choose the beginning and end of the time segments in order to enhance the statistical outcome. In this paper we show that the mathematical concept of autocorrelation functions (correlator functions) can be applied to find evidence of circadian rhythm in noisy data.

**Materials and Methods**

*Databases*

All results in text and figures are presented as mean ± s.e.m. Two databases were used, the Cardiac Arrhythmia Suppression Trial (CAST) which enrolled myocardial infarction patients from 1987 ~1989 (8), and Innovative Stratification of Arrhythmic Risk (ISAR), which enrolled myocardial infarction patients from 1996 ~2000 (2). The CAST data were studied because frequent VPCs (an average of 6 or more per hour on Holter monitoring) was a requisite to enrolment. The ISAR data were studied because they represented a population receiving myocardial infarction treatment considered optimal today. Qualifying pre-drug treatment Holter records of 751 patients in the CAST database were considered. Qualifying records were defined as being greater than 18 hours in duration, containing more than 15 VPCs, and excluded atrial fibrillation and paced rhythms (9). Of these records, there were 684 records with at least 25 usable VPCs, where usable is defined in the next section. All CAST analyses are based on these 684 records. Patients from the ISAR database (n=1455) with at least 25 usable VPCs in their Holter records were selected. All ISAR analyses are based on these 327 records.

*Calculation of HRT parameters TS and TO*

TS and TO were calculated in the standard way (1). Briefly, the 16 RR intervals flanking each VPC were extracted as a row of values and aligned such that the VPC coupling interval was in the third column. Each column was averaged to give an average HRT response profile. TS was defined as the steepest slope over 5 consecutive RR intervals of the rising portion of the HRT response profile. TO was defined as the difference between the two RR intervals preceding the VPC coupling interval and the two succeeding the compensatory pause, expressed as a percentage of the former. Criteria for selection of VPCs were: i) VPCs had to be classified as such, ii) VPCs had to be preceded by and followed by a minimum of 12 normal beats, where normal was defined as RR interval <2000 ms and classified as normal, iii) VPC coupling interval was at least 100 ms shorter than the RR interval preceding the VPC, iv) compensatory pause was at least 100 ms longer than the RR interval preceding the VPC. These VPCs were called usable VPCs. Only hours with at least 5 usable VPCs were included in circadian analysis of HRT parameters. Hours are denoted in 24 hour clock time, e.g., the interval 3 to 4 PM was called hour 15. TS >2.5 ms/beat and TO < 0% were considered normal.

*Calculation of Correlator Function Specific to this Study*

We describe the method for calculating correlator functions using the example of TS, although the same method was applied to other parameters. For a single patient we calculate the covariance between the values of TS at different times:

$$C_{TS}(\Delta hr) = \frac{1}{hrs} \sum_{hr} (TS(hr) - \overline{TS}) (TS(hr + \Delta hr) - \overline{TS})$$

where *hrs* is the number of hours for which there were at least 5 VPCs in both hour *hr* and in hour (*hr + Δhr*). We did not wrap our times around at 24 hours: *hr + Δhr* is always a later time in the patient's record than *hr*. Only integer values of *Δhr* were considered. TS for a patient at hour *hr* is denoted TS(*hr*), and the mean of the TS over all the hours that contained at least 5 VPCs is denoted $\overline{TS}$. The covariance functions of the patients were averaged to produce a population correlator function:

$$\overline{C_{TS}(\Delta hr)} = \frac{1}{pts} \sum_{pt} C_{TS}(\Delta hr)$$

The errors on the correlator function were obtained by the bootstrap method (10,11): the calculation was repeated for 100 different resamplings of the full set of patients, yielding a mean and s.e.m. for each value of the correlator function. To evaluate whether a given correlator function $C[\Delta t]$ was significantly different from the no-rhythm value $C_0 \cong -(C[0]-C[1])/24$ (see Appendix), and therefore indicative of a circadian rhythm, we calculated a chi-squared to quantify its deviation,

$$\chi^2 = \sum_{\Delta t=1}^{23} \left( \frac{C[\Delta t]-C_0}{C_{sem}[\Delta t]} \right)^2$$

where $C_{sem}[\Delta t]$ is the standard error of $C[\Delta t]$, and we do not include $\Delta t=0$ in the sum. If this chi-squared (with 23 degrees of freedom) is significantly large, then we reject the hypothesis that the data is pure noise, and infer a circadian rhythm.

The criteria for demonstrating a circadian rhythm were that the correlator function had to be significantly different from this null-hypothesis ("no-rhythm") value, and had to have a U-shape consisting of positive values at $\Delta t \cong 0$ and 24 hours and negative values at some intermediate $\Delta t$. To understand why we use this criterion, imagine a patient whose TS shows a circadian rhythm such that his TS is lower in the daytime, and higher during sleep. The correlator function at $\Delta t = 24$ will be positive, because it is the deviation of TS at a given time from the mean, multiplied by the deviation of TS from the mean 24 hours later. E.g., (TS(t)-TSmean) x (TS(t+24)-TSmean) is positive because if we choose t during sleep, say at 3 AM, TS is greater than the mean, as will be TS at 3 AM the next day, so (TS(t)-TSmean) and (TS(t+24)-TSmean) are both positive, and their product is positive. If we choose t during the day, say at noon, TS is below the mean, as will be TS at noon the next day, so (TS(t)-TSmean) and (TS(t+24)-TSmean) will both be negative, but their product will again be positive. On the other hand, the correlator at $\Delta t=12$ will likely be negative, because if TS is greater than the mean at a given time, it will tend to be less than the mean twelve hours later: (TS(t)-TSmean) > 0 and (TS(t+12)-TSmean) < 0, and their product will be negative. The same is true if TS is smaller than the mean at a given time. It will tend to be greater than the mean twelve hours later, so (TS(t)-TSmean) < 0 and (TS(t+12)-TSmean) > 0, and their product will be negative. More technical rationale for using these criteria is elaborated upon in the Appendix.

**Results**

*Demographics*

Patient characteristics for the full CAST and ISAR databases are described elsewhere (9,2). For the subset of 684 CAST patients in this study, mean age was 61.0 ± 0.4, 18% were female, 22% were diabetic, 42% had a history of myocardial infarction prior to the index infarction, left ventricular ejection fraction was 37.0 ± 0.4%, and 31% had received thrombolysis. The percentage of patients on diuretics was 33%, beta blockers 34%, aspirin 71%. There were 359 and 141 patients in whom TS and TO were both normal or both abnormal, respectively. For the subset of 327 ISAR patients in this study, mean age was 62 ± 0.6, 23% were female, 22% were diabetic, 22% had a history of myocardial infarction prior to the index infarction, left ventricular ejection fraction was 49.0 ± 0.9%. Seven % had received thrombolysis, and 86%, percutaneous coronary intervention. The percentage of patients on ACE inhibitors was 89%, diuretics 44%, beta blockers 89%, aspirin 97%, statins 83%. There were 150 and 47 patients in whom TS and TO were both normal or both abnormal, respectively.

*Circadian rhythm in heart rate and VPC frequency*

Heart rate was assessed by the RR interval (time interval between electrocardiographic QRS peaks). The mean RR interval over the day was shorter in CAST than in ISAR (808 ± 58.8 *vs.* 904 ± 8.3 ms), but the circadian pattern in RR interval was obvious on inspection for both datasets (Figure 1, top left). RR intervals preceding the VPCs were shorter than the RR interval for normal sinus beats (difference 4.4 ms, p=0.03 in CAST, difference 21.8 ms, p<0.0001 in ISAR), especially during the night hours. The RR interval correlator functions of the two datasets both had a U shape that crossed the correlation = 0 line at $\Delta$hr=5, reached a minimum near $\Delta$hr=11, and returned to positive values as $\Delta$ hr approached 24 (Figure 1 top right). Comparing the measured correlators with the no-rhythm value (-72 for CAST, -80 for ISAR) gave a very large chi-squared/degree of freedom (178 for CAST, 60 for ISAR, both p<0.0001) indicating presence of circadian rhythm.

There were a total of 208,774 usable VPCs in CAST. VPC count/patient was 305.2 ±

10.0, median 225, range 24~2058. VPC count/patient/hour was 13.8 ± 0.4, median 10, range 0~86. In the ISAR data, there were a total of 50,560 usable VPCs. VPC count/patient was 185.9 ± 11.4, median 124, range 20~1295. VPC count/patient/hour was 9.1 ± 0.5, median 6.1, range 0~56. Thus, usable VPC count over the day and per hour was approximately 60% higher in CAST. The pattern plots of VPC frequency showed some indication of circadian variation, much weaker than in the RR interval data, for both CAST and ISAR (Figure 1 bottom left). However, the correlator functions had the U-shape that is characteristic of circadian rhythm (Figure 1 bottom right). Comparing the measured correlators with the no-rhythm value (-3 for CAST, -2 for ISAR) gave a very large chi-squared/degree of freedom (482 for CAST, 138 for ISAR, both p<0.0001), indicating presence of circadian rhythm.

*Circadian rhythm in HRT parameters*

In CAST, the pattern plot of TS but not TO was suggestive of circadian variability (Figure 2, left panels). TS was higher than the average value of 5.21 in the hours 20~7, and lower in most of the remaining hours. TO fluctuated about the average value of -1.27. However the correlators gave a much clearer signal: both the TS and TO correlator functions had a U-shape, indicating circadian rhythm (Figure 2, right panels). The deviation from the no-rhythm value was highly significant, with a chi-squared/degree of freedom of 10.0 and 11.3 (both p<0.0001) respectively for TS and TO. In ISAR, both the pattern plots of TS and TO had large error bars and appeared to be random (Figure 3, left panels). Their correlator functions had an approximately U-shape, but the large error bars at late $\Delta t$ made it difficult to conclude that the last few values were above the no-rhythm line (Figure 3, right panels). Nevertheless, the deviation from the no-rhythm line was highly significant, with a chi-squared/degree of freedom of 3.4 and 3.0 (both p<0.0001) respectively for TS and TO, giving quantitative evidence of circadian rhythmicity.

*Heart rate and HRT*

TS has been shown to depend on heart rate both across (12,13) and within individuals

(14). Therefore, we calculated the linear regression between the 24 mean values of RR and HRT parameters. We found a positive correlation between RR and TS in CAST ($r^2$=0.64, $p<10^{-5}$) but not in ISAR ($r^2$=0.04, p=0.3). We found no correlation between the hourly averages of RR and TO in CAST ($r^2$=0.05, p=0.3) or ISAR ($r^2$=0.1, p=0.1). We next calculated TS values normalized to a heart rate of 75 as suggested by Hallstrom *et al* (7) for each hour in CAST. The correlator function for normalized and non-normalized TS was essentially the same. It crossed the correlation = 0 line at $\Delta$hr=4, had a minimum at $\Delta$hr=12, and showed the same rise to positive values at the end.

A cross-correlator function can be calculated similarly to the correlator function to test for phase differences between parameters. Because a study by Burgess *et al* (15) found that respiratory sinus arrhythmia, a measure of parasympathetic activity, began to increase 2 hours before sleep onset, and changes in it lagged changes in core body temperature by 5 hours, we tested for phase differences in our study parameters in patients with normal HRT in CAST. The peak correlation for both TS and TO with RR occurred at $\Delta$hr=0, negating the likelihood of any phase difference between HRT and RR. The TO-RR correlator showed an inverted U shape, indicating that TO tends to be low when RR is high, and vice versa.

*Influence of beta-blocker usage*

In CAST, the 273 patients on beta blockers had slower heart rate, fewer VPCs, better TS (6.1 ± 0.5 vs 4.8 ± 0.3), and better TO (-1.6 ± 0.2 vs -1.1 ± 0.1), compared to the 564 patients not receiving beta-blockers. The pattern plots of RR interval, VPC count, TS, and TO in patients receiving beta-blockers did not suggest that beta-blocker usage blunted circadian variability, except that the difference between maximum and minimum RR interval was smaller for patients on beta-blockers. Patients on beta blockers showed a very slightly more U-like shape of their TS correlator function (lower nadir, earlier return to positive values). TO correlator functions were similar for patients on and not on beta-blockers.

**Discussion**

We had expected that HRT would show a clear circadian pattern because of its close relation to heart rate variability, a phenomenon known to exhibit a circadian pattern (16-18). HRT parameters are weakly but significantly correlated with heart rate variability parameters (4,13,19-21), and in statistical analyses that include both HRT and heart rate variability as risk predictors, the two are rarely independent of each other (1,2,4). However, we were only able to show persuasive evidence of a circadian rhythm in HRT by calculating correlator functions. This suggests that a circadian pattern cannot be used to select optimal hours for measuring HRT in patients with a history of myocardial infarction. We propose two reasons for the similarity between day and night HRT values. One is the lower VPC count during sleep due to the sleep process (22,23) and lower heart rate (24), which leads to less reliable HRT values during night hours. The second is that in some patients, VPCs are associated with increased sympathetic tone (25). If in a subset of patients, VPCs tend to accompany transient increases in sympathetic tone, then, a study of HRT necessarily selects periods during the night when sympathetic tone is elevated.  Such periods would not be representative of average night-time autonomic tone in those patients. E.g., muscle sympathetic nerve activity matching daytime values are known to occur in REM sleep (26). In this study, we observed that night-time RR interval preceding VPCs were shorter than RR intervals of the surrounding hour, although they were not as short as daytime values. Comparing HRT induced by paced beats (12) with HRT following spontaneous VPCs over the night would be useful in testing this hypothesis.

*Previous studies of circadian rhythm in HRT*

Hallstrom *et al* (7) used CAST data to present pattern plots of TS. They presented TS values from 12 two hour bins, both as raw values and as values normalized to a heart rate of 75. The plot of the raw values indicated that TS was low during the day when heart rate was high (hour 8~18), and high during the night when heart rate was low (hour 0~8) similar to our results. Normalized TS showed a reduced circadian rhythm. They concluded that the circadian pattern in TS was due to dependence of TS on heart rate. The circadian pattern of TS was more distinct than ours, possibly due to different selection criteria of usable VPCs. The authors stated that TO was constant over the day. More recently, they presented pattern plots of heart rate- and VPC count- normalized TS for patients who died and those who didn't in CAST. A circadian pattern was clear in patients who died, and statistical analysis showed TS recorded from hours 8~18 was predictive of death (27). Another study analyzed data from 46 patients with ischemic heart disease and at least 100 VPCs on their Holter records. The authors compared median HRT values between three 4 hour time periods selected to represent morning, afternoon and night (28). They found that afternoon TS was lower than night TS, and that TO had no circadian variability. In contrast to these studies which found a circadian rhythm only in TS, we were able to demonstrate a circadian rhythm in TS normalized by heart rate and in TO as well. A fourth study reportedly found that HRT was more marked during sleep in young healthy people (29).

*CAST- ISAR difference*

Why was circadian rhythm in HRT harder to demonstrate in ISAR than in CAST? Although there were twice as many patients in CAST leading to smaller error bars, the main reason is likely to be the 60% higher VPC count in CAST. For each patient, the analysis ignored hours in which <5 VPCs were present, so the ISAR data had more gaps. In addition, although the conventionally recommended minimum VPC count for HRT calculation is 5 (3), TS values do not stabilize until VPC count is much higher (R. Schneider, unpublished data). Therefore, CAST should reflect changes in autonomic tone more accurately than ISAR.

In alternative efforts to demonstrate circadian rhythm in ISAR pattern plots, we tried a bin size of two hours to increase the number of VPCs per bin, but this did not produce better pattern plots (data not shown). A recent study showed that only patients in whom TS and TO were both normal were able to manifest a change in TS before imminent ventricular tachyarrhythmia (30). One conclusion of that study was that patients with poor HRT have less variability of their HRT values. We therefore also tried limiting patients to those in whom both HRT parameters were normal. However, the resulting correlator functions were essentially the same, but with larger error bars (data not shown). This is interesting because CAST patients who remained alive were shown to have less circadian variability of TS than patients who died, in the Hallstrom study (27) mentioned above.

A second difference between CAST and ISAR that may possibly affect the circadian pattern of HRT is that the Holter electrocardiograms in CAST were recorded at a range of times after myocardial infarction, including at home, while in ISAR, they were recorded within two weeks of myocardial infarction in a hospital setting. TO values may improve with time after myocardial infarction (31), although there is disagreement (32). If TO improves with time, some of the TO values may be better in CAST because of the later date of the Holter recording relative to their AMI. However, the proportion of patients with abnormal TO was very similar (23% in CAST and 19% in ISAR). It is also likely that in a hospital setting, patients sleep poorly at night and are less active during the day, leading to a more disturbed circadian pattern than that of patients whose Holter is recorded at home. The clean circadian pattern observed in heart rate in the ISAR data does not imply uninterrupted sleep, because the circadian pattern in heart rate is not altered by forced overnight wakefulness (15).

A third possible reason for the difference between CAST and ISAR is that only 34% of CAST patients were on beta-blockers in contrast to 89% in ISAR. I.e., it may be that beta-blockade reduces circadian variability. A study by Pitzalis suggests that beta-blockade reduces heart rate mostly during the day (33), thereby reducing day-night heart rate difference. We compared patients on and not on beta-blockers in CAST. The difference between maximum and minimum RR interval of the pattern plot was smaller in the patients on beta-blockers. However, beta-blocker usage did not appear to blunt the circadian patterns of TS and TO, suggesting that beta-blocker usage per se was not responsible for the stronger circadian rhythm pattern in CAST. Instead, beta-blockade likely affects HRT values indirectly by reducing VPC frequency and heart rate. It should be noted that acute beta-blockade did not affect HRT values in an induced HRT study (34), and that the effectiveness of HRT as a risk predictor has remained high despite increases in beta-blocker usage over the years. Finally, beta-blocker therapy has been shown to increase TS along with left ventricular ejection fraction in patients with congestive heart failure (35), demonstrating that beta-blockers can influence HRT.

*Question of heart rate dependence of HRT*

Does HRT have an intrinsic circadian rhythm, or does it merely mirror circadian changes in heart rate? Although HRT is regularly described as depending on heart rate because of the correlation between them, correlation is not causation, and it seems rational to assume that both heart rate and HRT ultimately follow the circadian pattern of autonomic tone set by higher neural centres. Vagal outflow is the predominant determinant of heart rate, and appears to be the predominant determinant of HRT (6,34) as well. It is difficult to think of a physiological dependence of HRT on heart rate that is unrelated to their mutual dependence on baseline autonomic tone. There is, however, a dependence of HRT on heart rate because of the way TS is calculated (7). We therefore studied the correlator function of TS normalized by heart rate for CAST, but this did not change the correlator function. Having said that, elimination of circadian rhythm in HRT after normalization would not have proven lack of an intrinsic circadian rhythm in HRT. E.g., if body temperature had a circadian pattern roughly similar to heart rate, normalizing HRT by body temperature would also abolish the circadian pattern of HRT, without proving dependence or causation between them. In contrast to TS, TO values are normalized by heart rate by definition. Our results thus make clear that both parameters of HRT have a circadian rhythm beyond a first order correlation with heart rate.

*Correlator function as a method for studying circadian rhythm*

We found the correlator function to be a very sensitive detector of presence of circadian rhythm, in variables where the rhythm was nearly invisible in the simple pattern plot. We believe the correlator function is powerful because of its insensitivity to differences in baseline values between patients (the average is subtracted off separately for each patient) and because of its insensitivity to phase differences in the circadian patterns between patients (it depends only on time differences within each patient's record, dispensing with the need to align data by moment of waking, say). It can be used for studying records with unknown start times, and for studying phase relations between two parameters, as we did in this study. The main disadvantage of the correlator function is that it does not specify what the actual circadian pattern is. Even when the correlator function indicates a strong 24 hour cycle, it is possible that the phase and detailed shape of the cycle are variable between patients.

Cosinor analysis is another quantitative and fine time scale method for studying circadian rhythm (18,36). It utilizes regression of 24 hour data onto a trigonometric function. We did not apply cosinor analysis because the RR interval pattern plot which showed the most obvious circadian rhythm was not sinusoidal, and because gaps in hours with usable VPCs made curve fitting difficult in individual patients. It would also have suffered from the same sensitivity to difference in sleep-wake phase and baseline values between individuals as pattern plots.

We note that regardless of the method used, using a 24 hour record to assess a rhythm with a hypothesized 24 hour period is mathematically dubious, and is valid only under the assumption that the 24 hour record would approximately repeat itself if measured for a longer period. However, 24 hour recordings have long been the norm in clinical cardiology. As we discuss in the Appendix, the correlator function becomes a much more powerful diagnostic indicator of circadian rhythms if longer recordings are used; even extending the recording by a few extra hours greatly increases its discriminatory power.

**Appendix**

The criteria for demonstrating a circadian rhythm were that the correlator function had to be significantly different from the null-hypothesis ("no-rhythm") value, and had to have a U-shape consisting of positive values at $\Delta t \cong 0$ and 24 hours and negative values at some intermediate $\Delta t$. We illustrate the derivation of these criteria using data simulated from artificial circadian rhythms. The dataset of a variable $V$ measured for times $t$ (typically 0 to 23) from 100 patients ($i$=0 to 99), was generated according to:

$$V_i(t) = A\ P(t) + \varsigma_s$$

where $P(t)$ is the circadian rhythm, $A$ is the amplitude, and noise $\varsigma_s$ is a Gaussian-distributed random number with standard deviation $s$. We kept the circadian amplitude $A$ and the function $P(t)$ fixed from patient to patient for simplicity of presentation. To calculate the correlator function, with bootstrapping to obtain reliable $1\sigma$ error bars, we followed the procedure as described in Methods for TS.

1. *Correlator functions of periodic rhythms* (Figure 4 top two rows)

A periodic rhythm produces a correlator function with positive values at $\Delta t \cong 0$ and $\Delta t \cong$ period (and multiples of the period), and negative values significantly below the null-hypothesis value (see next section) at some intermediate times. This criterion reflects the fact that a circadian pattern will deviate from the mean in the same direction at times separated by 24 hours, but in opposite directions at smaller time separations (typically $\Delta hr \cong 12$ hours for a smooth circadian pattern). In the top row of Figure 4, we show a cosine function circadian rhythm $P(t)$ with a 24 hour period, and the corresponding correlation function obtained after combining $P(t)$ (amplitude $A$=2) with random noise of standard deviation $s$=2. Positive correlations occur at $\Delta t \cong 0$ and $\Delta t \cong$ period, and negative correlations occur in between. In the middle row of Figure 4, we show that a circadian rhythm with a more irregular underlying pattern still gives a positive correlator at $\Delta t \cong 24$ hrs, and a dip to negative values at intermediate times.

2. *Calculating the null-hypothesis (no-rhythm) value*

The top right panel of Figure 4 also shows a correlation function for pure noise of standard deviation $s$=2 (white squares). At $\Delta t$=0 we find Corr[0] $\cong s^2$. At later times one might expect the correlator to drop to zero because noise at different times is uncorrelated, but in fact it drops to a small negative value, $C_0 = -s^2/24$. Roughly speaking, this is due to the finite number, i.e., 24, of data points used, and the fact that if the value of $V$ at some time $t$ is above the mean, then on average the other values have to be slightly below the mean and vice versa. In our ISAR and CAST data we do not know $s^2$ *a priori*, but we can estimate it from the data. As the first two rows of Figure 4 show, as long as any underlying circadian rhythm is reasonably smooth, the noise always contributes $s^2$ to Corr[0] and a negligible amount to Corr[1], so we can estimate $s^2 \cong$ Corr[0]-Corr[1], so the no-rhythm line (the null hypothesis value) is at about -(Corr[0]-Corr[1])/$T$ for a recording of length $T$ hours. This estimate of the null-hypothesis value is shown as a dashed line in Figures 2 and 3. In Figure 1, the no-rhythm lines were not shown, because they were indistinguishable from the correlation = 0 line for the y axis scale used.

3. *The effect of recording length* (Figure 4 bottom row)

In Figure 4, as well as in our data (Figures 2 and 3), the error bars on the correlators become larger as Δt increases. This is because at Δt=1, there are 23 possible contributions <$V$(0)$V$(1)>, ..., <$V$(22)$V$(23)>, to the correlator function, but this number drops as Δt increases to 23, where there is only one contribution, <$V$(0)$V$(23)>, and a smaller number of measurements leads to a larger standard error of estimate. This increase in the error bars as Δt approaches the duration $T$ of the recording means that in a 24 hour recording, it can be hard to show that the correlator returns to positive values at Δt≅24, which is a signature of a circadian rhythm. This is illustrated in the bottom row of Figure 4. The left panel shows the correlator for simulated data with a cosine rhythm with lower amplitude $A$=1, while the noise has the same standard deviation $s$=2 as in the rest of Figure 4. The return to positive values at Δt ≅ 24 is obscured by noise. In the bottom right panel we show the correlator obtained from a simulation of longer duration $T$=36 hours. The positivity of the correlator at Δt ≅ 24 becomes much clearer.

**References**

1. Schmidt G, Malik M, Barthel P, Schneider R, Ulm K, Rolnitzky L, Camm AJ, Bigger JT Jr, Schömig A: Heart-rate turbulence after ventricular premature beats as a predictor of mortality after acute myocardial infarction. *Lancet.* 1999;353:1390-1396.

2. Barthel P, Schneider R, Bauer A, Ulm K, Schmitt C, Schömig A, Schmidt G: Risk stratification after acute myocardial infarction by heart rate turbulence. *Circulation.* 2003;108:1221-1226.

3. Watanabe MA, Schmidt G: Heart rate turbulence: a 5 year review. *Heart Rhythm.* 2004;1:732-738.

4. Ghuran A, Reid F, La Rovere MT, Schmidt G, Bigger JT, Camm AJ, Schwartz PJ, Malik M: Heart rate turbulence-based predictors of fatal and nonfatal cardiac arrest (The Autonomic Tone and Reflexes After Myocardial Infarction substudy). *Am J Cardiol.* 2002;89:184-90.

5. Mäkikallio TH, Barthel P, Schneider R, Bauer A, Tapanainen JM, Tulppo MP, Schmidt G, Huikuri HV: Prediction of sudden cardiac death after acute myocardial infarction: role of Holter monitoring in the modern treatment era. *European Heart J.* 2005;26:762-769.

6. Marine JE, Watanabe MA, Smith TW, Monahan KM: Effect of atropine on heart rate turbulence. *Am J Cardiol.* 2002;89:767-9.

7. Hallstrom AP, Stein PK, Schneider R, Hodges M, Schmidt G, Ulm K: Structural relationships between measures based on heart beat intervals: Potential for improved risk assessment. *IEEE Transactions on Biomed Engineering.* 2004;51:1414-1420.

8. Echt DS, Liebson PR, Mitchell B, Peters RW, Obias-Manno D, Barker AH, Arensberg D, Baker A, Friedman L, Greene HL, Huther ML, Richardson DW, and the CAST Investigators: Mortality and morbidity in patients receiving encainaide, flecainide, or placebo. The Cardiac Arrhythmia Suppression Trial. *New Eng J Med.* 1991;324:781-788.

9. Stein PK, Domitrovich PP, Kleiger RE, Schechtman KB, and Rottman JN: Clinical and demographic determinants of heart rate variability in patients post myocardial infarction: Insights from the cardiac arrhythmia suppression trial (CAST). *Clin Cardiol.* 2000;23:187-194.

10. Diaconis P and Efron B: Computer intensive methods in statistics. A gentle introduction to the bootstrap. *Scientific American.* 1983:248:116-30.

11. Davison AC, Hinkley DV: A comprehensive survey of applications with theory of the bootstrap, including the double bootstrap. *Bootstrap Methods and Their Application.* Cambridge: Cambridge University Press, 1997.

12. Watanabe MA, Marine JE, Sheldon M, Josephson ME: Effects of ventricular premature stimulus coupling interval on blood pressure and heart rate turbulence. *Circulation.* 2000;106:325-330.

13. Cygankiewicz I, Wranicz JK, Bolinska H, Zaslonka J, Zareba W: Relationship between heart rate

turbulence and heart rate, heart rate variability, and number of ventricular premature beats in coronary patients. *J Cardiovasc Electrophysiol.* 2004;15:731-737.

14. Bauer A, Malik M, Barthel P, Schneider R, Watanabe MA, Camm J, Schömig A, Schmidt G: Turbulence dynamics: An independent predictor of late mortality after acute myocardial infarction. *Internat J Cardiol.* 2006;102:42-47.

15. Burgess HJ, Trinder J, Kim Y, Luke D: Sleep and circadian influences on cardiac autonomic nervous system activity. *Am J Physiol.* 1997;273:H1761-1768.

16. Ewing DJ, Clarke BF: Diagnosis and management of diabetic autonomic neuropathy. *Brit Med J.* 1982;285:916-918.

17. Massin MM, Maeyns K, Withofs N, Ravet F, Gerard P: Circadian rhythm of heart rate and heart rate variability. *Arch Dis Child.* 2000;83:179-182.

18. Guo Y-F, Stein PK: Circadian rhythm in the cardiovascular system: Considerations in non-invasive electrophysiology. *Cardiac Electrophysiology Review.* 2002;6:267-272.

19. Turkmen A, Ider YZ, Sade E, Aytemir K, Oto A & Kucukdeveci F: Relation between heart rate turbulence and heart rate variability spectral components. Proceedings IEEE EMBS International Conference 2001.

20. Koyama J, Watanabe J, Yamada A, Koseki Y, Konno Y, Toda S, Shinozaki T, Miura M, Fukuchi & Ninomiya M: Evaluation of HRT as a new prognostic marker in patients with chronic heart failure. *Circulation J.* 2002;66,902-907.

21. Lindgren KS, Mäkikallio TH, Seppänen T, Raatikainen MJP, Castellanos A, Myerburg RJ, Huikuri HV: Heart rate turbulence after ventricular and atrial premature beats in subjects without structural heart disease. *J Cardiovasc Electrophysiol.* 2003;14:447-452.

22. Lown B, Tykocinski M, Garfein A, Brooks P: Sleep and ventricular premature beats. *Circulation.* 1973;48:691-701.

23. Pickering TG, Goulding L, Cobern BA: Diurnal variation in ventricular ectopic beats and heart rate. *Cardiovasc Med.* 1977;2:1013-1027.

24. Winkle RA: The relationship between ventricular ectopic beat frequency and heart rate. *Circulation.* 1982;66:439-446.

25. Lown B, Verrier RL, Rabinowitz SH: Neural and psychologic mechanisms and the problem of sudden cardiac death. *Am J Cardiol.* 1977;39:890-902.

26. Hornyak M, Cejnar M, Elam M, Matousek M, Wallin BG: Sympathetic muscle nerve activity during sleep in man. *Brain.*1991;114:1281-95.

27. Hallstrom AP, Stein PK, Schneider R, Hodges M, Schmidt G, Ulm K: Characteristics of heart beat intervals and prediction of death. *International J Cardiol.* 2005;100:37-45.

28. Cygankiewicz I, Wranicz JK, Bolinska H, Zaslonka J, Zareba W: Circadian changes in heart rate turbulence parameters. *J Electrocardiology.* 2004;37:297-303.

29. Guzik P, Schmidt G: A phenomenon of heart-rate turbulence, its evaluation, and prognostic value. *Card Electrophysiol Review.* 2002;6:256-261.

30. Watanabe MA: Heart rate turbulence slope reduction in imminent ventricular tachyarrhythmia and its implications. *J Cardiovasc Electrophysiol.* 2006;17:735-740.

31. Jokinen V, Tapanainen JM, Seppänen T, Huikuri HV: Temporal changes and prognostic significance of measures of heart rate dynamics after acute myocardial infarction in the beta-blocking era. *Am J Cardiol.* 2003;92:907-912.

32. Ortak J, Weitz G, Wiegand UKH, Bode F, Eberhardt F, Katus HA, Richardt G, Schunkert H, Bonnemeier H: Changes of heart rate, heart rate variability and heart rate turbulence throughout the course of reperfused myocardial infarction. *Pacing Clin Electrophysiol.* 2005;28,227-232.

33. Pitzalis MV, Mastropasqua F, Massari F, Totaro P, Scrutinio D, Rizzon P. Sleep suppression of ventricular arrhythmias: a predictor of beta-blocker efficacy. Eur Heart J 1996;17:917-925.

34. Lin LY, Lai LP, Lin JL, Du C, Shau W, Chan H, Tseng Y, Huang SKS: Tight mechanism correlation between heart rate turbulence and baroreflex sensitivity: sequential autonomic blockade analysis. *J Cardiovasc Electrophysiol.* 2002;13:427-431.

35. Lin LY, Hwang JJ, Lai LP, Chan HL, Du CC, Tseng YZ, Lin JL. Restoration of heart rate

turbulence by titrated beta-blocker therapy in patients with advanced congestive heart failure: positive correlation with enhanced vagal modulation of heart rate. *J Cardiovasc Electrophysiol.* 2004;15:752-6.

36. Halberg F, Johnson EA, Nelson W, Runge W, Sothern R: Autorhythmometry procedures for physiologic self-measurements and their analysis. *Physiol Teacher.* 1973;1:1-11.

**Figure Legends**

**Figure 1**. Circadian rhythm in heart rate and VPC frequency for CAST and ISAR patients.

Top left: Pattern plot of RR interval. Top right: Correlator function for RR interval. Bottom left: Pattern plot of VPC count/hr. Bottom right: Correlator function for VPC count/hr. Filled circles: CAST (n=684); unfilled circles: ISAR (n=327). Plots display mean ± s.e.m.

**Figure 2**. Circadian rhythm in heart rate turbulence parameters for CAST patients (n=684).

(a) Pattern plot of TS. (b) Correlator function for TS. (c) Pattern plot of TO. (d) Correlator function for TO. Plots display mean ± s.e.m. Dashed line shows expected correlator function at $\Delta t>0$ if there were no circadian rhythm. The TS pattern plot displays a circadian rhythm, the TO pattern plot does not. However, the correlator functions of both strongly suggest presence of a circadian rhythm.

**Figure 3**. Circadian rhythm in heart rate turbulence parameters for ISAR patients (n=327).

(a) Pattern plot of TS. (b) Correlator function for TS. (c) Pattern plot of TO. (d) Correlator function for TO. Plots display mean ± s.e.m. Dashed line shows expected correlator function if there were no circadian rhythm. Neither of the pattern plots display a circadian rhythm. The correlator functions suggest presence of a circadian rhythm, but the size of the error bars increases with $\Delta t$.

**Figure 4**. Simulated correlator functions. Plots display mean ± s.e.m. **First two rows:** Correlation functions for artificially generated 24-hour data samples, showing underlying circadian pattern *P(t)* (left panels) and resultant correlator function (right panels). In each case the rhythm is multiplied by amplitude factor *A*=2 and combined with noise of standard deviation *s*=2. Top row: cosine circadian rhythm *P(t)*=cos(2π *t*/24) (filled symbols) and no rhythm *P(t)*=0 (unfilled symbols). Middle row: less regular circadian rhythm. Dashed lines shows expected correlator function at $\Delta t>0$ if there were no circadian rhythm. **Bottom row:** Correlator functions for artificially generated data consisting of 24-hour cosine circadian rhythm of amplitude *A*=1 combined with noise of standard deviation *s*=2. Left panel: 24 hour recording time. Right panel: 36 hour recording time. The positive deviations from the no rhythm line at $\Delta t \cong 24$ become clearer when recording durations are extended.

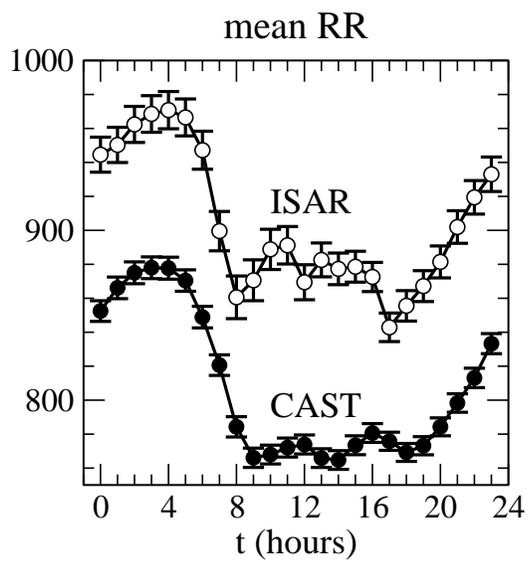

**mean RR**

**RR correlator**

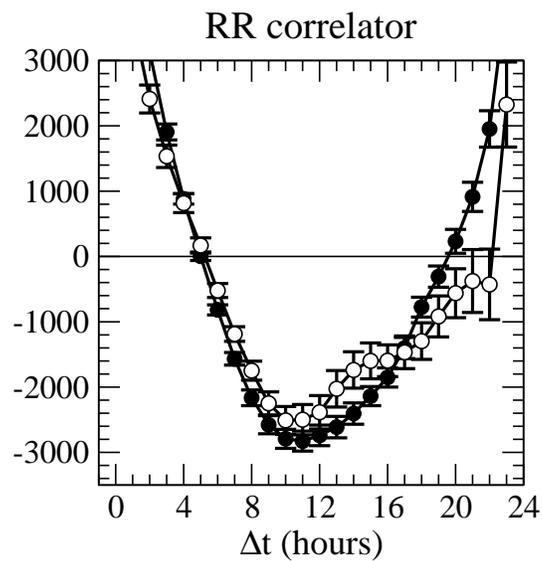

**mean VPCs/hr**

**VPCs/hr correlator**

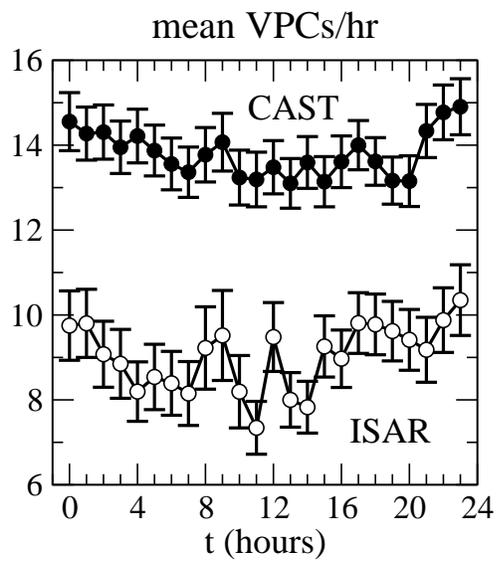

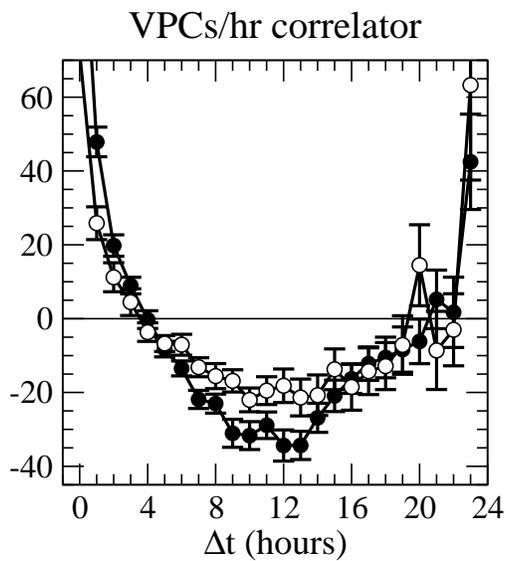

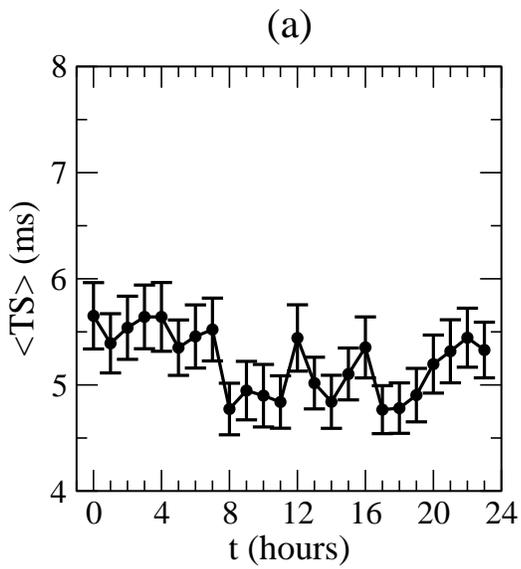

(a)

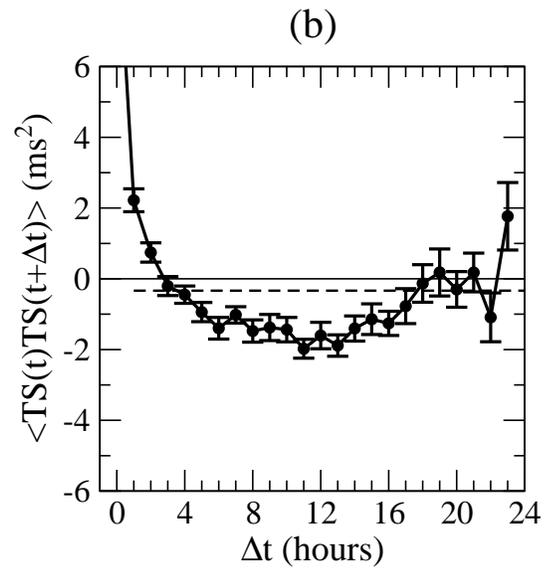

(b)

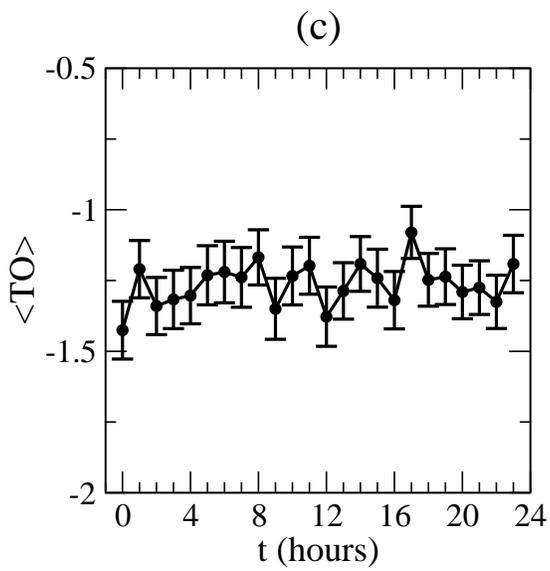

(c)

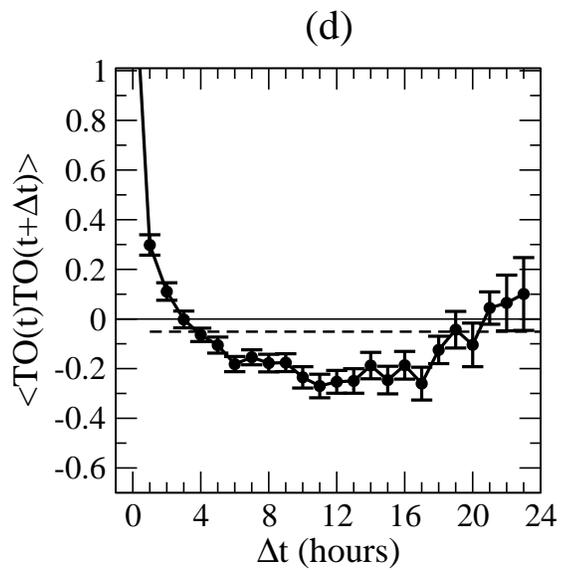

(d)

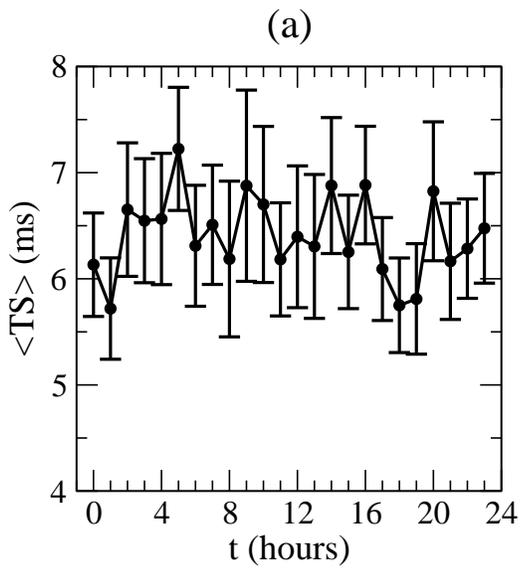
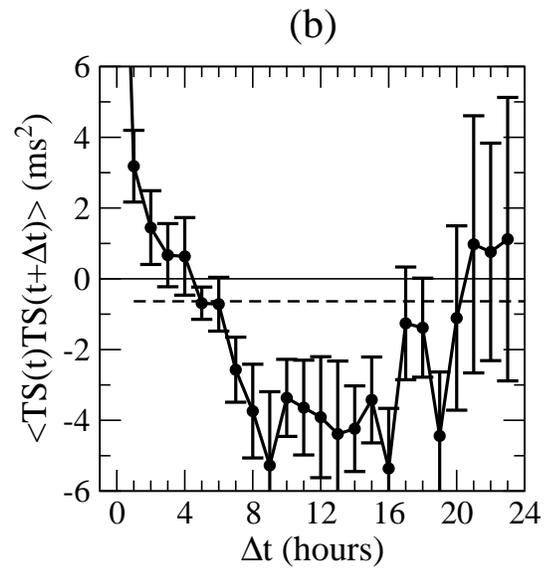
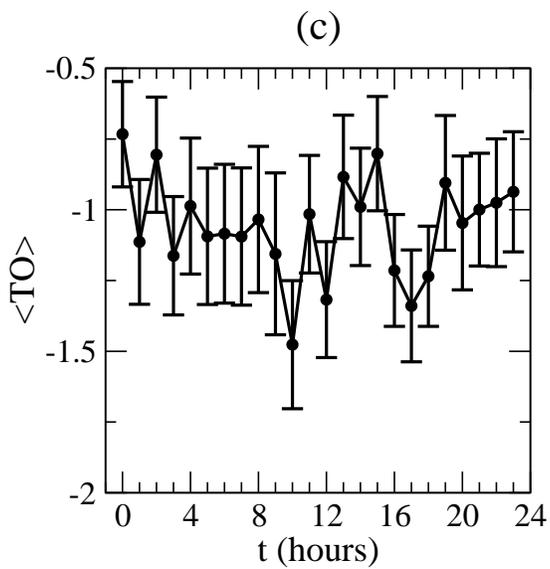
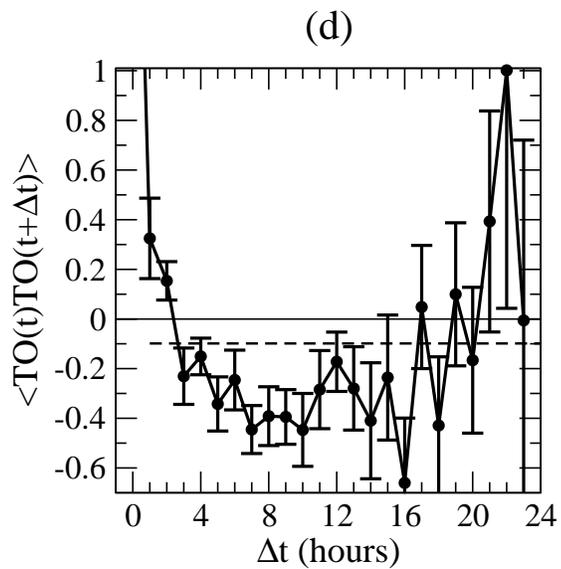

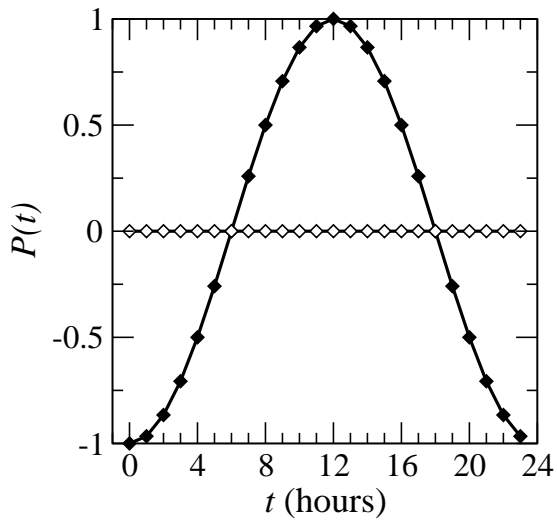
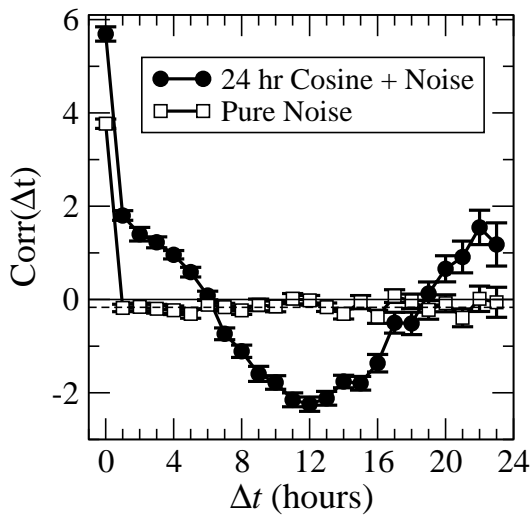
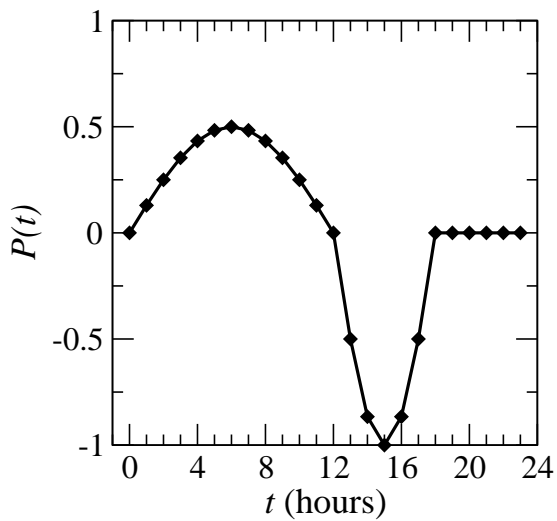
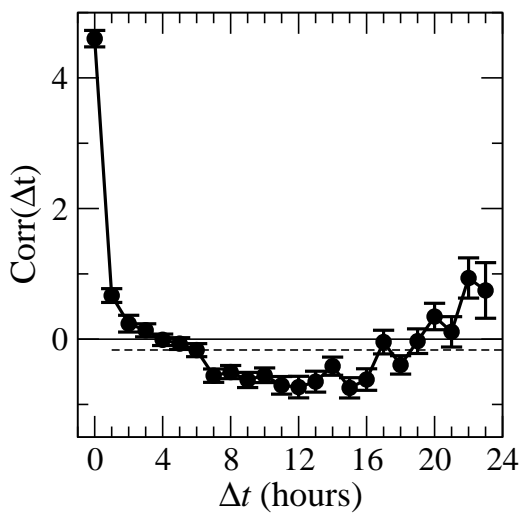
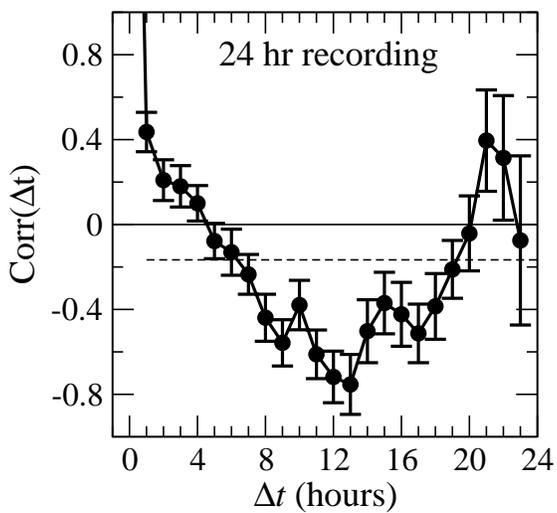
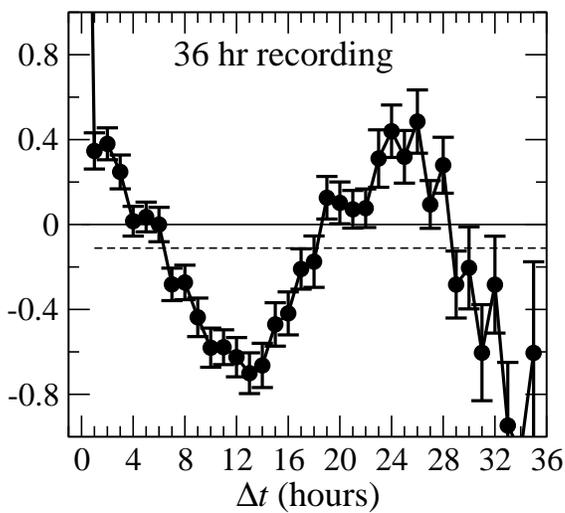